\documentstyle[11pt,newpasp,twoside,epsf]{article}
\def\edcomment#1{\iffalse\marginpar{\raggedright\sl#1\/}\else\relax\fi}
\reversemarginpar

\begin{document}
\title{
Spectroscopy of BL Lac Objects: new redshifts and mis-identified sources.}
 \author{N. Carangelo$^{1,4}$}
\affil{$^1$ Universit\`a di Milano-Bicocca, 
 P.za della Scienza 3, 20126 Milano, I 
\\E-mail:nicoletta.carangelo@mib.infn.it}
\author{R. Falomo$^2$, J. Kotilainen$^3$ , A. Treves$^4$ and M.-H. Ulrich$^5$}
\affil{$^2$ Osservatorio Astronomico di Padova;
$^3$ Tuorla Observatory;\\
$^4$ Universit\`a dell'Insubria;
$^5$ European Southern Observatory}

\begin{abstract}
We are carrying out a program of high signal to noise optical spectroscopy of BL Lacs with  unknown or tentative redshift. Here we report some preliminary results.  New redshifts are measured for PKS0754+100 (z=0.266) and 1ES0715-259 (z=0.464). From  lineless spectra of PG1553+113 and PKS1722+119 we set a lower limit  of z$>$0.3 for both sources. In two cases (UM493 and 1620+103) stellar spectra indicate a wrong classification.

\end{abstract}

\section{Introduction and Observations}

At variance with other classes of AGN BL Lac Objects are characterized by quasi featureless optical spectra. This often hinders the determination of the redshift (and therefore of the distance) of these sources. In the V\'eron C\'etty and V\'eron catalog (2001) of AGN  there are $\sim$600 objects classified as BL Lacs and only for half of them the redshift is known. In order to contribute to the measurement of this fundamental parameter we are carrying out a systematic study of the optical spectra of objects with unknown  or tentative redshift. We present here preliminary results from two campaigns at the European Southern Observatory (ESO) telescope performed in 2001 July and 2002 January.

The observations were gathered using the 3.6m telescope equipped with EFOSC2.  Standard data reduction was performed using different packages in IRAF in order to obtain 1-dimensional wavelength and flux calibrated extracted spectra.
\section{Results}
\subsection{New redshifts}

{\bf PKS 0754+100} is a highly polarized and variable source that was discovered by Tapia et al. (1977) and classified as a blazar by Angel \& Stockman (1980). A first tentative redshift proposed by Persic and Salucci (1986, z=0.66) was put in doubt after the detection of the host galaxy by Abraham et al. (1991) and by Falomo (1996), because at this redshift the host would be extremely luminous (M$_{R}\simeq$-25).

Our spectrum (see fig. 1) shows clearly 
 two emission lines at $\lambda$=4717.2 {\AA} (EW=0.9 {\AA}) and $\lambda$=6340.4 {\AA}(EW=1.3 {\AA}).
 Identification of these features with [OII](3727 {\AA}) and [OIII] (5007 {\AA}) yields a redshift z=0.266$\pm$0.001. This is consistent with the tentative redshift proposed by  Falomo and Ulrich (2000, z$\simeq$0.28).
At z=0.266 the host galaxy luminosity becomes M$_{R}$=-22.9. We note that at the same redshift there is a companion galaxy, 13 $''$ NE, corresponding to a projected distance of $\sim$ 70 kpc (Pesce et al. 1995).\\

{\bf 1ES 0715-259} belongs to the {\em Einstein} Slew Survey (Elvis et al. 1992) and was classified as a BL Lac Object by Perlman et al. (1996). They  assumed that a VLA radio source could be identified with the corresponding Slew Survey source if it fell within 80$''$ of X-ray position. However in this case they reported an offset in the radio-X position of $>$200 $''$. The VLA radio map of 1ES 0715-259 (Perlman et al. 1996)
shows a clear FR II morphology.
The MMT spectrum obtained by Perlman et al.(1996) is featureless which motivated the BL Lacs classification.
The value of the ratio F$_{x}$/F$_{r}$ (logF$_{x}$/F$_{r}$=-5.18) places 1ES 0715-259 near the borderline between RBL and XBL objects. There are no data of polarimetry for this source.  
HST (Urry et al. 2000) clearly resolved 1ES0715-259 into a point source surrounded by a small, rather elongated host galaxy.

We took an optical spectrum of the counterpart of the radio source. This
shows  clearly strong emission lines at $\lambda$=7114.5 (EW=85.3 {\AA}), $\lambda$=7262.4 {\AA} (EW= 14.3 {\AA}) and $\lambda$=7332.0 {\AA} (EW=46 {\AA}) (see figure 1). Identification  respectively with the H$_{\beta}$, [OIII] 4959  {\AA } and [OIII] 5007 {\AA} yields a redshift z=0.464$\pm$0.001. The H$_{\beta}$ luminosity ( L$_{H_{\beta}}$=2.5 $\times$ 10$^{43}$ erg s$^{-1}$) is typical of radio loud quasar. On the basis of our optical spectroscopy and the radio properties we believe that this source is not a BL Lac but a radio loud quasar. The mis-identification  by Perlman et al. (1996) is likely due to the very poor quality of their optical spectrum.\\

\begin{figure}
\plottwo{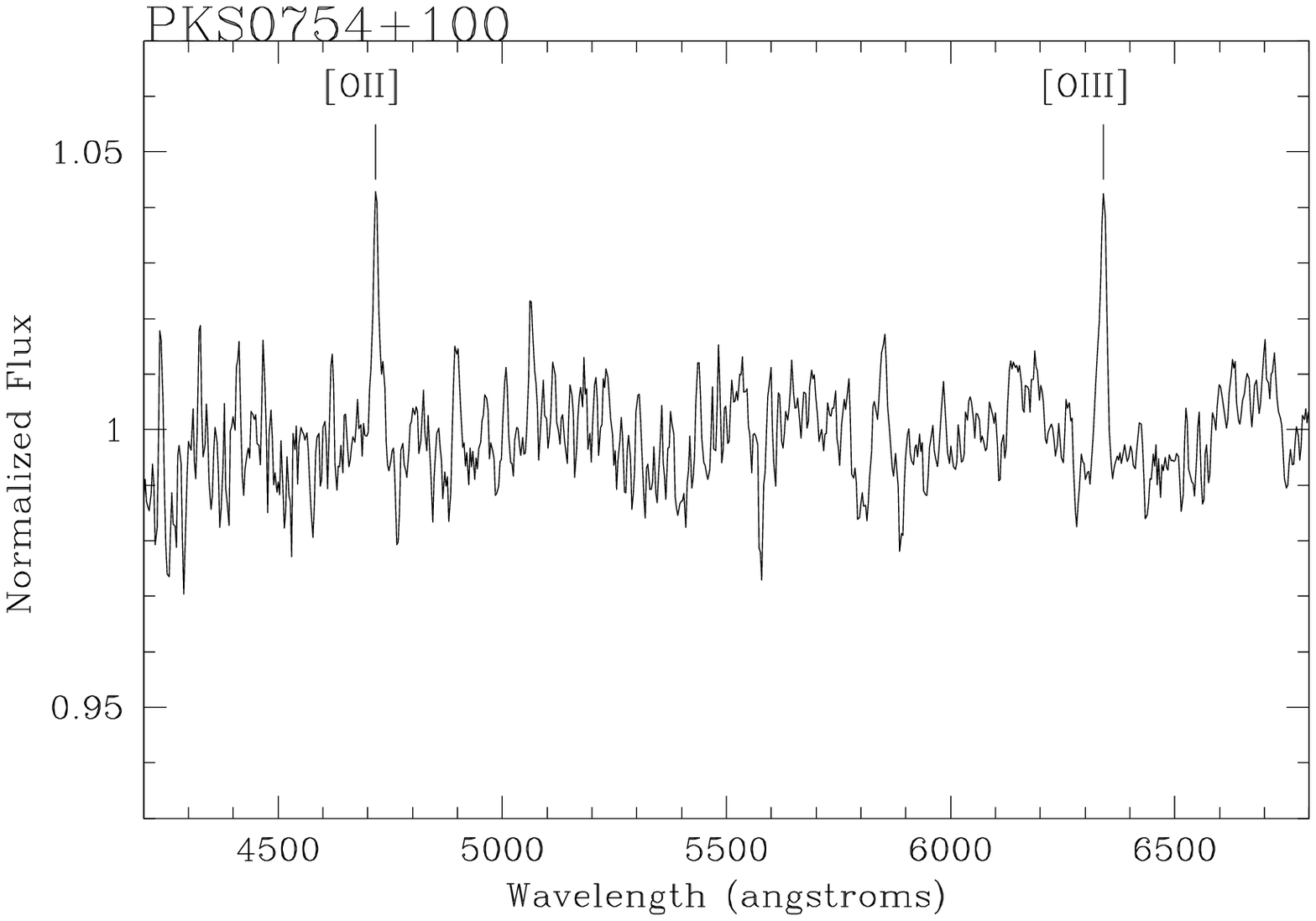}{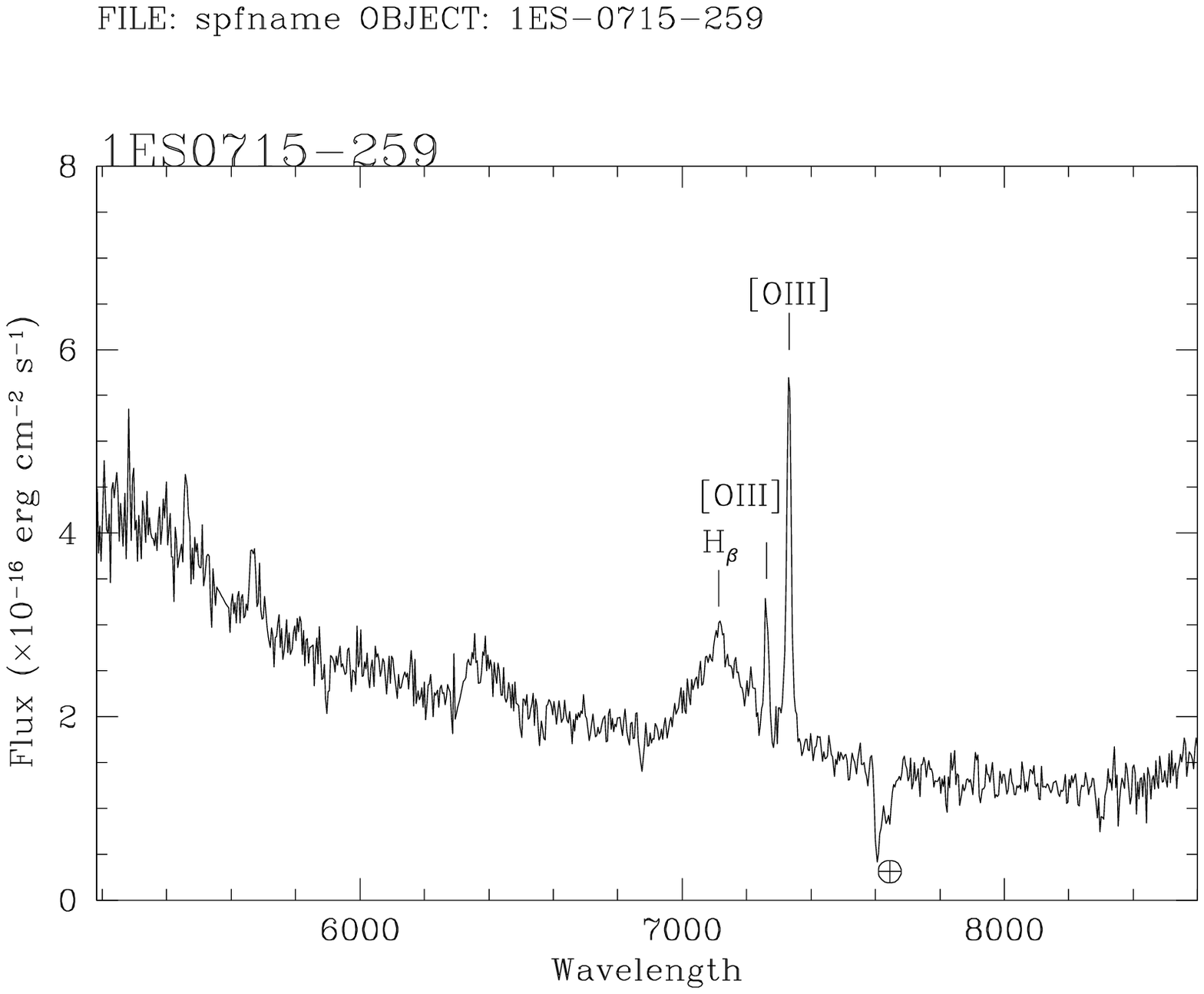}
\caption{Left: The normalized spectrum of PKS0754+100. The two emission lines at z=0.266 are marked. Right: Spectrum of 1ES0715-259 showing clearly strong emission lines at z=0.464 of a typical QSO.}
\end{figure}
 
\begin{figure}  
\plottwo{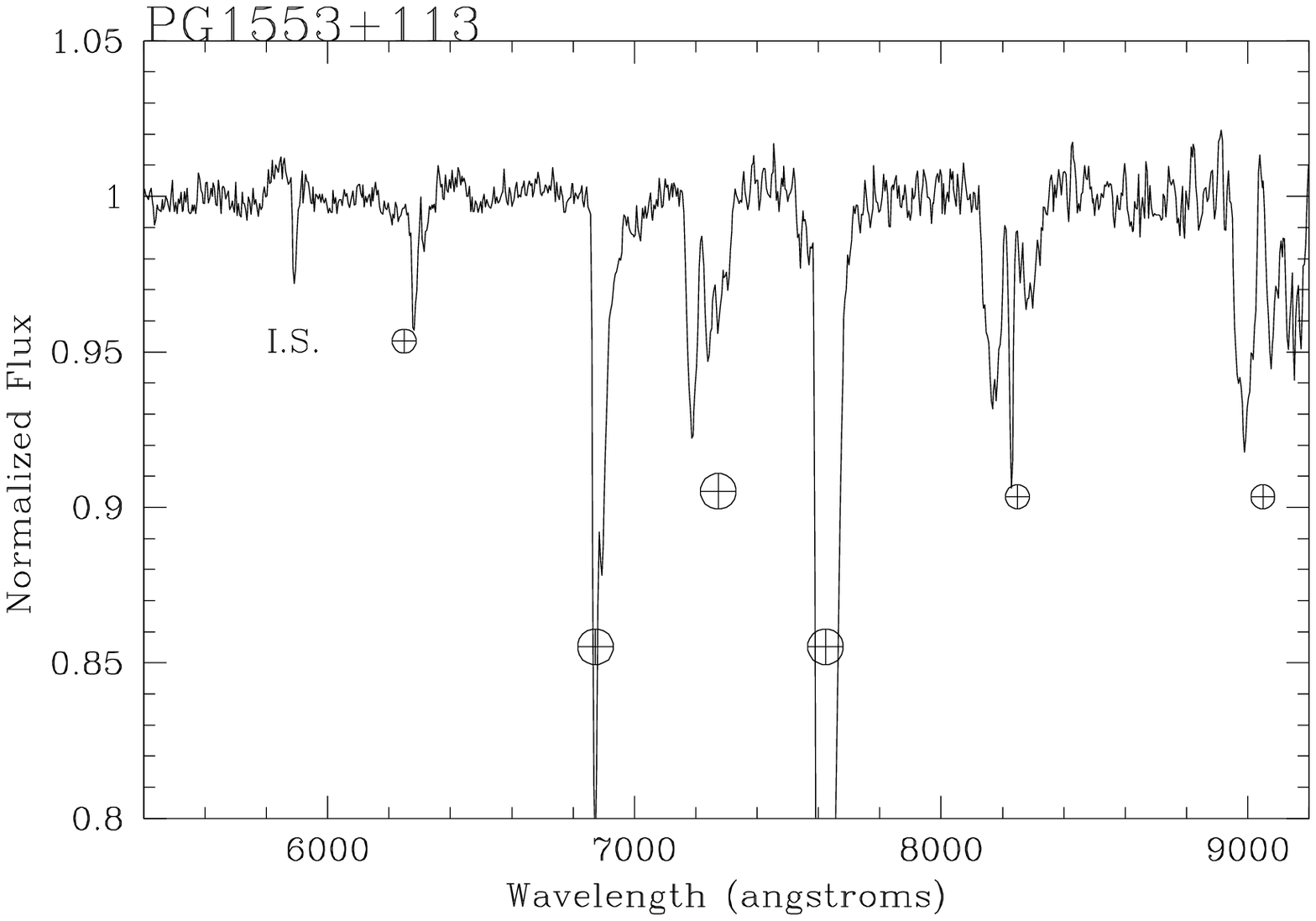}{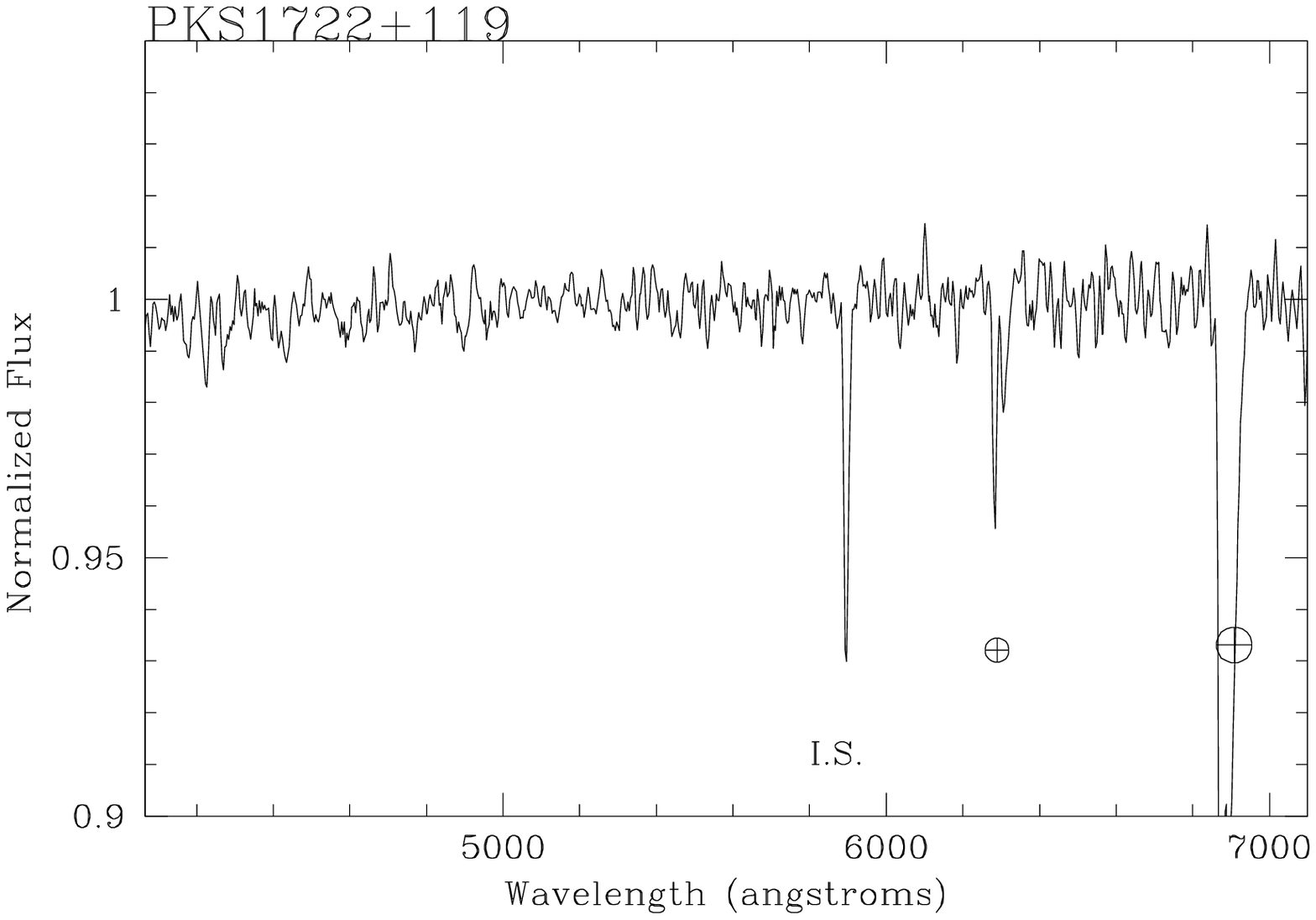}
\caption{The normalized spectra of PG1553+113 (left) and PKS1722+119 (right). The spectra have S/N=200 and are lineless (EW$<$0.1 {\AA}).}
\end{figure}

\subsection{Lineless objects}

For some relatively bright objects (V$\sim$ 15-16) despite the high signal to noise (S/N$\sim$200) the spectra appear lineless. We  report here two examples: PG1553+ 113 and PKS1722+119 (see figure 2).\\ 

{\bf PG 1553+113} is a very bright optically selected source (Green et al. 1986) classified as a BL Lacs on the basis of the featureless spectrum (Miller \& Green 1983) and the large optical variability ($\Delta$m$\simeq$ 2 mag, Miller et al. 1988). Falomo \& Treves (1990) showed that the emission feature in the UV region  identified as Lyman-$\alpha$ at z=0.36 by Miller \& Green (1983) is spurious. Falomo \& Treves (1990) 
did not find absorption features with an EW upper limit of 1 {\AA}. 

Our high signal to noise optical spectrum of the object confirms the absence of lines with a much lower of EW ($\sim$0.1 {\AA}). The redshift of this source remains unknown. Assuming that it is hosted  by a galaxy typical of BL Lacs (M$_{R}$=-23.7 R$_{e}$=10 kpc, Urry et al. 2000) from our spectrum we set a lower limit of z$>$0.3 consistent with the fact that the source is unresolved in HST images (Urry et al. 2000). \\

{\bf PKS1722+119} is a highly polarized X-ray selected BL Lac Object originally discovered by Uhuru. Griffith et al. (1989) claimed the observation of an absorption feature near 6000 {\AA} which could be Na I (5892 {\AA}) redshifted at z=0.018. This was disproved by Brissenden et al. (1990) who obtained a completely lineless spectrum, consistently with more recently results by  V\`eron C\`etty and V\`eron 1993 and by Falomo et al. 1993. 

The spectrum presented here (see fig. 2) is lineless and we can set a lower limit of z$>$0.3. Also this source is unresolved by HST (Urry et al. 2000).

\subsection{Mis-identified sources}
From the spectra examined so far we found two cases UM493 and 1620+103 where we observe the typical stellar absorption  lines of the Balmer series clearly indicating a wrong classification. Probably the mis-identifications (e.g.V\`eron-C\`etty \& V\`eron  2001, Hewitt \& Burbidge 1993) were due to the poor informations on these sources, classification being based only on modest optical spectroscopy. Our high signal-to noise spectra clarify their nature and we can definitively classify these sources as stars of spectral type A.

\end{document}